# Mathematical Modeling of Extinction of Inhomogeneous Populations

G.P. Karev[1*] and I. Kareva[2,3]

[1] National Center for Biotechnology Information, National Institutes of Health - Bldg. 38A, 8600 Rockville Pike, Bethesda, MD 20894, USA. Email: karev@ncbi.nlm.nih.gov.

[2] Floating Hospital for Children at Tufts Medical Center, Boston, MA, 02111, USA.

[3] Simon A. Levin Mathematical, Computational and Modeling Sciences Center, Engineering Center A, Arizona State University, Tempe, AZ, 85287, USA. Email: ikareva@asu.edu.

*Correspondence should be sent to:

Georgy P. Karev, PhD

National Center for Biotechnology Information,

National Institutes of Health,

Bldg. 38A, 8600 Rockville Pike, Bethesda, MD 20894, USA.

Email: karev@ncbi.nlm.nih.gov

**Abstract**

Mathematical models of population extinction have a variety of applications in such areas as ecology, paleontology and conservation biology. Here we propose and investigate two types of sub-exponential models of population extinction. Unlike the more traditional exponential models, the life duration of sub-exponential models is finite. In the first model, the population is assumed to be composed clones that are independent from each other. In the second model, we assume that the size of the population as a whole decreases according to the sub-exponential equation. We then investigate the "unobserved heterogeneity", i.e. the underlying inhomogeneous population model, and calculate the distribution of frequencies of clones for both models. We show that the dynamics of frequencies in the first model is governed by the principle of minimum of Tsallis information loss. In the second model, the notion of "internal population time" is proposed; with respect to the internal time, the dynamics of frequencies is governed by the principle of minimum of Shannon information loss. The results of this analysis show that the principle of minimum of information loss is the underlying law for the evolution of a broad class of models of population extinction. Finally, we propose a possible application of this modeling framework to mechanisms underlying time perception.

**Introduction**

Questions of primary importance in modeling biological populations most frequently pertain to describing laws that govern population growth and development. Questions that focus on population decline and extinction as a primary focus of investigation appear to attract less attention despite their potential importance in the areas of ecology, paleontology and environment preservation. Furthermore, one could ostensibly look at individual death from this point of view as well, where any individual and even any single organ can be viewed as a complex community of many different cell populations that undergo extinction with the death of the individual. For instance, human brain is one of the largest biological populations, consisting of approximately $10^{11}$ neurons and $10^{14}$ synapses. Looking at extinction of this population from the point of view of the framework that will be presented here can provide unexpected and thought provoking predictions. An investigation of the process of population extinction with the help of even seemingly uncomplicated conceptual models can thus be of general interest.

The simplest model of population decrease and extinction is exponential of the form $\frac{dN}{dt} = -kN$, where *k* is some positive constant and $N(t)$ measures the current value of some property relevant to the process; typically $N(t)$ is the total population size. Radioactive decay can be described by this equation with high accuracy but in application to biological populations, the exponential model may be oversimplified. The model describes a population, which formally is immortal, and ignores an important fact that all real populations are polymorphic, and hence different parts of the population decrease at different rates.

Non-homogeneous models of population decline address this issue, since in these types of models, death rate of the entire population depends on the distribution of properties in the population, and so different subpopulations decay at different rates. The problem of interest in this work is the evolution of the distribution of properties and of the related statistical features, such as mean, variance, and entropy for given initial distribution. Important examples come from forest ecology (Karev, 2003) and epidemic models (Dodson, 2008).

Here we consider non-exponential models of extinction of polymorphic populations, which show some new interesting phenomena compared to exponential models. Non-exponential power models of the form $\frac{dx}{dt} = kx^p$ were suggested by E. Szathmary and M. Smith (1997) in order to model pre-biological evolution of replicators. Three cases are distinguished: the exponential with $p=1$; the super-exponential, with $p>1$; and the sub-exponential, with $p<1$. The models imply "differential survival of the fittest", "survival of the common", and "survival of everybody" respectively. Well established examples of non-exponential growth apply to global demography (*p*=2,von Forster 1960; Kapitza 1996, 2010; Karev 2005) and some molecular replicator systems (*p*=1/2, von Kiedrowski 1993; see also Johansen, Sornette, 2001).

It is not always clear why non-exponential growth would be observed in reality. However, recently it was shown (Karev, Koonin 2013; Karev, 2014) that non-exponential power models of population growth can be understood within the frameworks of inhomogeneous population models.

An important property of sub-exponential models of extinction is that the lifetime of corresponding populations is finite. Here, we investigate two types of inhomogeneous power models of population extinction. In the first model we assume that the population is composed of independent clones, each of which decreases according to the sub-exponential equation; the problem of interest is the dynamics of total population size and distribution of densities of the clones. We show that the distribution dynamics follows the Principle of Minimum of the Tsallis information loss.

In the second model we assume that the total size of the inhomogeneous population decreases according to the sub-exponential equation; the problem of interest is the composition of the population and the dynamics of separate clones. We demonstrate within the frameworks of inhomogeneous population models the universal *frequency-dependent* representation of the power extinction models. We also show that the dynamics of clone distribution follows the Principle of Minimum of Shannon information loss. Finally, we propose that the population as a whole can possess "internal time", which tends to infinity as the "external", or chronological time tends to a finite time moment of population extinction. We conclude with a discussion of a possible interpretation of the properties of internal time, and how it can relate to time perception.

**Population of sub-exponentially decreasing clones**

In this section we consider the model of a population composed of individuals of distinct types; the extinction dynamics of *i*-th type of individuals is given by the equation:

$$\frac{dx_i}{dt} = -k_i x_i^p. \qquad (1.1)$$

We call a set of all individuals of a certain type a clone. Let us emphasize that the dynamics of each clone in this model does not depend on the total population or other clones.

The solution to equation (1.1) is given by

$$x_i(t) = x_i(0)(1 - x_i(0)^{p-1} k_i (1-p) t)_+^{\frac{1}{1-p}}. \qquad (1.2)$$

Here and henceforth the subscript "+" means a positive part of corresponding expression. Solution (1.2) can be conveniently written in the form

$$x_i(t) = x_i(0) \exp_p(-x_i(0)^{p-1} k_i t) \qquad (1.3)$$

where $\exp_q(x) \equiv (1+(1-q)x)_+^{1/(1-q)}$ is by definition the so called $q$-exponential function. The inverse of $\exp_q(x)$ is given by the $q$-logarithm function $\ln_q x = \dfrac{x^{1-q}-1}{1-q}$. These two functions tend to ordinary exponential and logarithm functions, respectively, as $q \to 1$ (see, e.g., Tsallis 2008, ch.3, for formulas and properties of the so-called $q$-calculus).

According to Equation (1.2), each clone has a finite life time. The $i$-th clone becomes completely extinct at the moment $T_i$ i.e. $x_i(T_i) = 0$, where extinction moment $T_i$ is given by

$$T_i = \frac{x_i(0)^{1-p}}{k_i(1-p)}. \qquad (1.4)$$

Notice that for exponential and super-exponential models (1.1) with $p \geq 1$, every clone has an indefinite life time. It is of course an unrealistic property, which is why we will focus primarily on sub-exponential extinction models.

Figure 1 shows typical dependence of the extinction moment $T_i$, defined in Equation (1.4), on the initial clone size and the value of the power $p$.

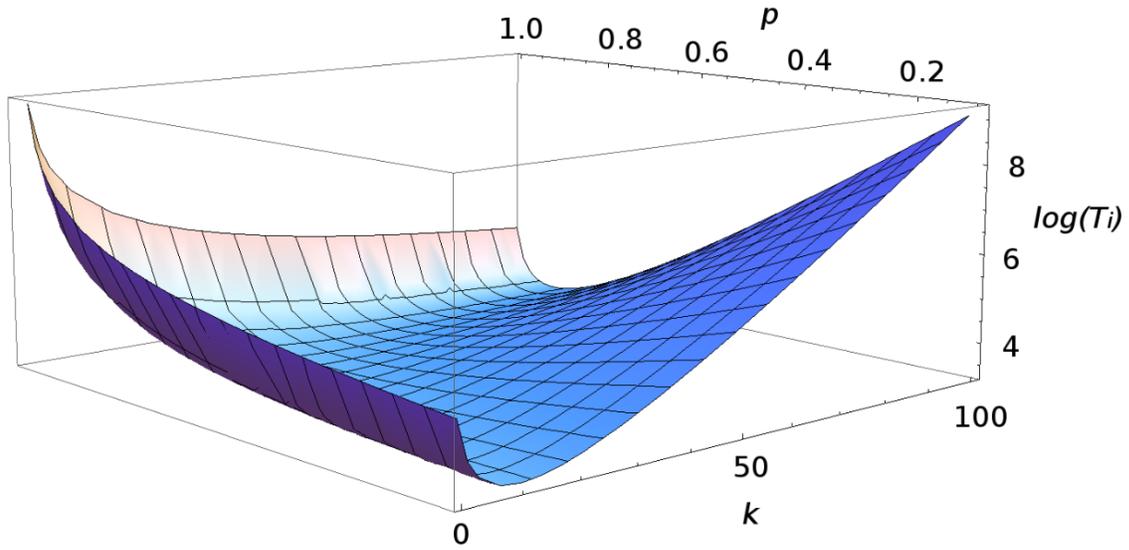

**Figure 1.** Dependence of logarithm of the extinction time $T_i$, defined in Equation (1.4), on the initial clone sizes $x_k = e^{-sk}, k = 1,\ldots 100$ and the power $p$; $s = -0.1$.

Using the notion of extinction time $T_i$, we can now express the current density of $i$-th clone as

$$x_i(t) = x_i(0)\left(1 - t/T_i\right)_+^{1/(1-p)} . \qquad (1.5)$$

The total population size is given by the formula

$$N(t) = \Sigma_i x_i(t) = N(0)\Sigma_i x_i(0)(1 - x_i(0)^{p-1} k_i (1-p)t)_+^{\frac{1}{1-p}} \qquad (1.6)$$

$$= N(0)\Sigma_i P_0(i)\left(1 - t/T_i\right)_+^{1/(1-p)}$$

The total population goes to extinction at the moment $T = max_i T_i$.

Typical dynamics of the total population size is shown in Figure 2.

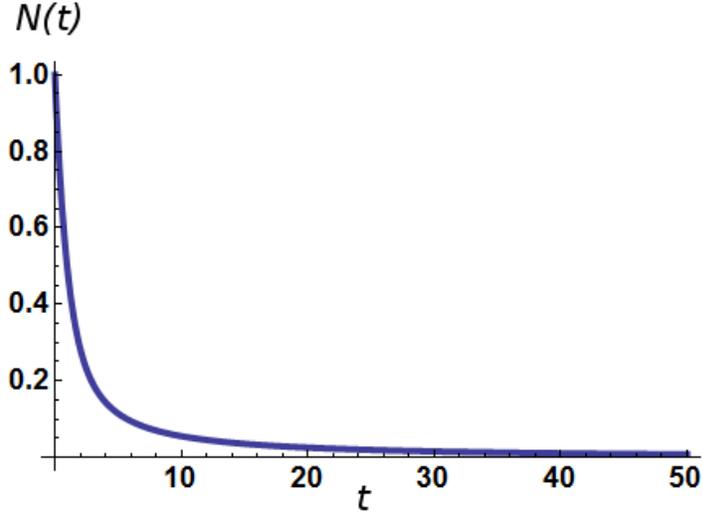

**Figure 2.** Typical dynamics over time of the total population size $N(t)$ defined according to Equation (1.6); $N(0)=1$, $P_0(i)=Ce^{-si}$, $i=1,\ldots100$, $C=1/\sum_{i=1}^{100}e^{-si}$ is the normalization constant; $p=0.9$, $s=0.1$.

The frequency of $i$-th individual is given by

$$P_t(i) = \frac{x_i(t)}{N(t)} = \frac{P_0(i)\left(1 - x_i(0)^{p-1} k_i (1-p)t\right)_+^{1/(1-p)}}{\Sigma_j P_0(j)(1 - x_j(0)^{p-1} k_j (1-p)t)_+^{\frac{1}{1-p}}}, \qquad (1.7)$$

which can also be written as

$$P_t(i) = P_0(i)\frac{(1 - t/T_i)_+^{1/(1-p)}}{\Sigma_j P_0(j)(1 - t/T_j)_+^{1/(1-p)}}. \qquad (1.8)$$

The dynamics of clone frequencies are shown in Figure 3.

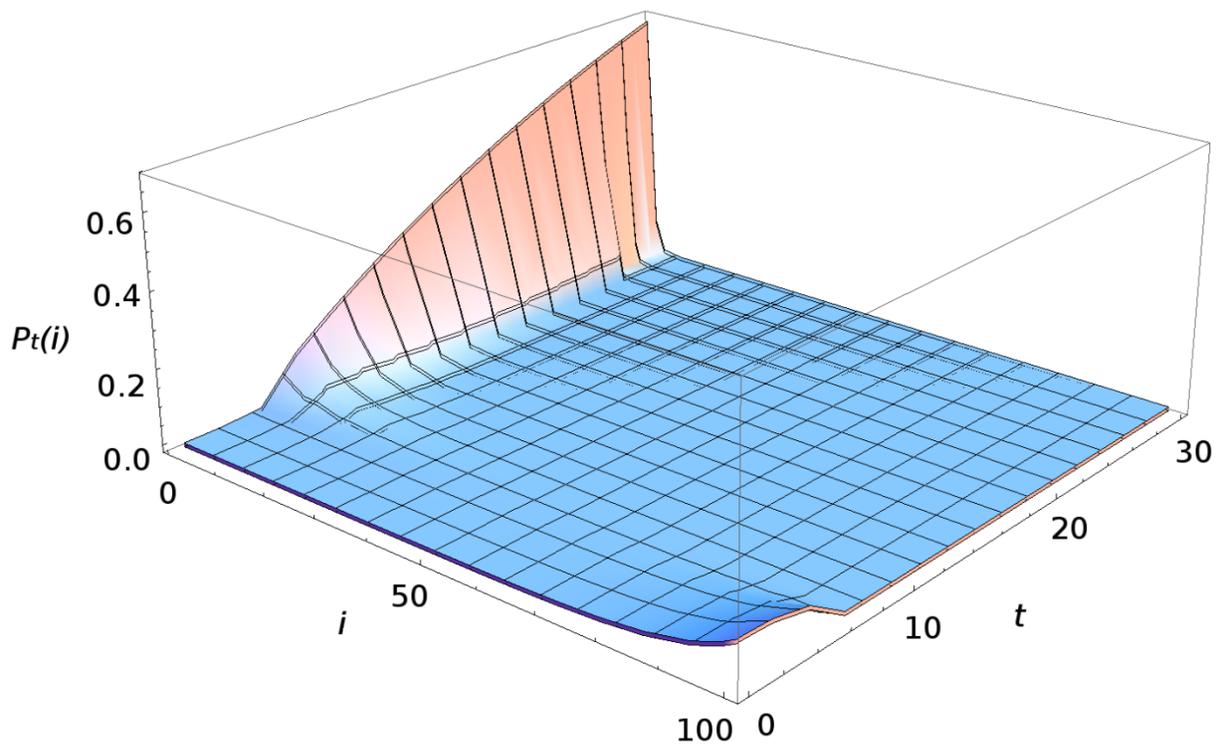

**Figure 3.** Dynamics of clone frequencies $P_t(i)$, defined in Equation (1.7); $P_0(i) = Ce^{-si}$, $i = 1,\ldots 100$; $p = 0.9, s = -1$.

### Dynamical principles of minimum of Tsallis information loss

We submit that the information measure for dynamical models and systems should be chosen in accordance with system dynamics. In the case of sub-exponentially decreasing populations, the distribution of individual frequencies (1.7) is the Tsallis distribution at each time moment, and accordingly, the Tsallis *q*-entropy is the appropriate information measure (for definitions see equation (2.1), (2.3) below).

The Tsallis relative *q*-entropy of a discrete probability distribution $\{m(i)\}$ given a reference distribution $\{r(i)\}$ is defined as:

$$I_q[m:r] = \frac{1}{q-1}\Sigma_i m(i)\left[\left(\frac{m(i)}{r(i)}\right)^{q-1} - 1\right] = \Sigma_i m(i) \log_q\left(\frac{m(i)}{r(i)}\right) \qquad (2.1)$$

(see Tsallis 2009, Borland et al, 1998 for definition, general properties and theorems). The polynomial divergence of the form (2.1) is known in statistics as "Cressie-Read divergence", see Cressie and Read 1984, Read and Cressie 1988; a brief useful survey of this and other non-classical entropies can be found in (Gorban et.al., 2010). The Tsallis relative *q*-entropy within the framework of population extinction model may be interpreted as a measure of information loss. The inference of *m* by minimizing $I_q[m:r]$ is known as the Principle of minimum of Tsallis relative *q*-entropy. The distribution that provides the minimum of $I_q[m:r]$ with respect to the constraint

$$\Sigma_i u(i) m(i)^q = <u>_q \qquad (2.2)$$

is the Tsallis distribution

$$m(i) = \frac{r(i)}{Z}\left[1 - (1-q)r(i)^{q-1}\beta u(i)\right]_+^{1/(1-q)} = \frac{r(i)}{Z}\exp_q\left(-r(i)^{q-1}u(i)\beta\right). \qquad (2.3)$$

Here Z is the normalization factor (the "*q*-partition function"):

$$Z(\beta) = \Sigma_i r(i)\left[1 - (1-q)r(i)^{q-1}u(i)\beta\right]^{\frac{1}{1-q}} = \Sigma_i r(i)\exp_q\left(-r(i)^{q-1}u(i)\beta\right). \qquad (2.4)$$

The Lagrange multiplier $\beta$ at a given constraint $<u>_q$ can be calculated from the equation

$$\frac{\partial}{\partial \beta}\ln_q Z = -<u>_q, \qquad (2.5)$$

where $\frac{\partial}{\partial \beta} ln_q Z = Z^{-q} \frac{\partial}{\partial \beta} Z$, as follows from the definition of $ln_q x$.

One can then calculate the minimum information loss as:

$$I_q[m:r] = -ln_q Z - \beta <u>_q = -ln_q Z + \beta \frac{\partial}{\partial \beta} ln_q Z. \qquad (2.6)$$

We can see that when $r(i) = P_0(i)$, $u(i) = N(0)^{p-1} k_i$, $q = p$, and $\beta = t$, the distribution (2.3) exactly coincides with the distribution (1.7), which can be written in the form:

$$P_t(i) = \frac{P_0(i)\left(1 - P_0(i)^{p-1} N(0)^{p-1} k_i (1-p)t\right)_+^{1/(1-p)}}{\Sigma_j P_0(j)\left(1 - P_0(j)^{p-1} N(0)^{p-1} k_j (1-p)t\right)_+^{1/(1-p)}} = \frac{P_0(i)}{Z} \exp_p\left(-P_0(i)^{p-1} N(0)^{p-1} k_i t\right) \qquad (2.7)$$

Notice that now the total population size can be written as $N(t) = N(0) Z(t)$.

Let us reformulate these results. We do not seek an unknown distribution that would minimize the relative Tsallis entropy subject to a particular set of constraints. Instead, we have the solution (1.2) of model (1.1), which produces distribution (2.7) at each time moment. With this distribution, we can compute at each moment the $p$-mean of the death rate, $\sum_i u_i P_t(i)^p \equiv <u>_p^t$. Importantly, we can compute this value knowing only the initial distribution $P_0(i)$ and initial population size $N(0)$. (Technically, in order to compute $<u>_p^t$, one can use the formula

$$<u>_p^t = -\frac{\partial}{\partial t} ln_p Z(t) = -Z^{-p} \frac{\partial}{\partial t} Z, \qquad (2.9)$$

where $Z(t)$ is defined by (2.4) with $q = p$).

Considering the computed value $<u>_q^t$ as a prescribed constraint, we can see that the distribution $P_t$ (2.7) provides minimum for the Tsallis information loss, $I_p[P_t : P_0]$. It means that within the frameworks of the model (1.1), the Principle of minimum of the Tsallis information loss is not a hypothesis, but a mathematical assertion, which follows from system dynamics. In other words, the dynamics of extinction model (1.1) is governed by the Principle of minimum of Tsallis information loss.

The following theorem holds, where we assume for simplicity that $N(0)=1$.

**Theorem 1**

*Distribution of individuals in a population of sub-exponentially decreasing clones provides the minimum of the Tsallis information loss $I_p[P_t:P_0]$ at each time moment t among all probability distributions that have the p-mean of the population death rate equal to $<k>_p^t$ at time moment t.*

The information loss $I_p[P_t:P_0]$ at $t$ moment can be calculated according to its definition (2.1) or using the formula

$$I_p[P_t:P_0]=-ln_p Z(t)-t<k>_p^t=-ln_p Z(t)+t\frac{\partial}{\partial t}ln_p Z(t). \qquad (2.10)$$

Figure 4 shows the dynamics of the Tsallis information loss with respect to different values of parameter *p,* when the initial distribution $P_0$ is geometric, $P_0(i)=Ce^{-i}, i=0,\ldots 100$ and *C* is the normalization constant.

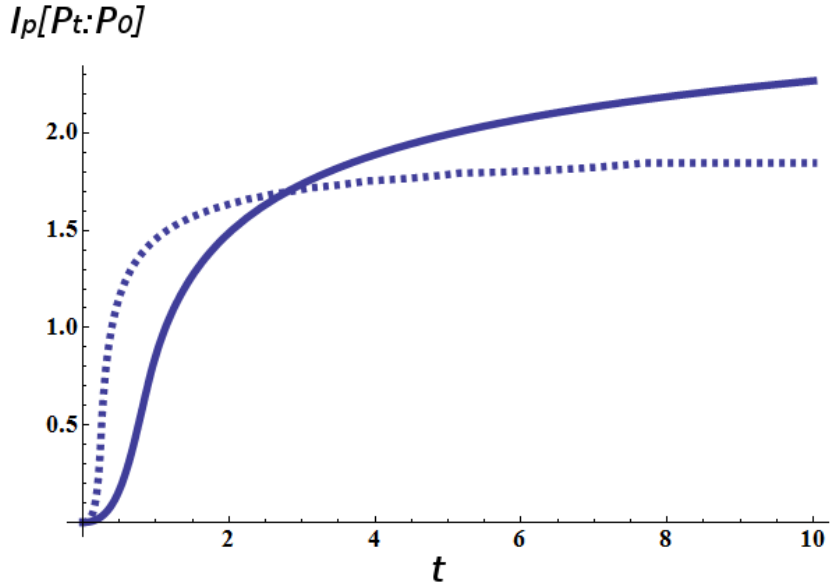

**Figure 4.** Dynamics of the Tsallis information loss $I_p[P_t : P_0]$ as defined in Equation (2.10) when $p = 0.5$ (dash) and $p = 0.7$ (solid).

An important and perhaps unexpected conclusion can be made now. We considered a population composed of independent clones, each of which decreases according to the power equation with the same power but with its own specific death rate. The clones do not depend either on each other or on the population as a whole. Therefore, it is difficult to expect the dynamics of such a population to admit some kind of "macro" description that ignores the "micro" dynamics of each independent clone. Nevertheless, such "macro" description does exist. We have shown that *given the initial distribution of clone frequencies in the population, the knowledge of only the p-mean of the death rates at any time moment implies complete knowledge of the population distribution at that moment due to the principle of minimum of Tsallis relative entropy (Tsallis information loss).* The corresponding Tsallis distribution coincides with the distribution (1.7), which follows from the "micro" dynamics of every clone. Let us emphasize that we do not claim a priori the principle of minimum of Tsallis *relative entropy*; we proved that this principle was fulfilled for power model (1.1) due to the "micro" dynamics of the population.

**Parametrically inhomogeneous models of population extinction**

In order to study the second type of inhomogeneous power models we need some preliminary mathematical results. A general mathematical framework for investigation of inhomogeneous population models was developed in (Karev, 2010 a, b). For clarity, the main results for a particular case of this type of a model, which is sufficient for our purposes, are given below.

Consider an inhomogeneous population composed of individuals with different death rates $a$; we refer to the set of all individuals with the given value of parameter $a$ as an $a$-clone. Let $l(t, a)$ be the size of $a$-clone at the moment $t$. We assume that the growth rate of each clone depends on the total population size $N(t)$. Dynamics of such a population can be described by the following model:

$$\frac{dl(t,a)}{dt} = -al(t,a)g(N), \qquad (3.1)$$

$$N(t) = \int_A l(t,a)da, \qquad (3.2)$$

where $g(N)$ is an appropriate function. Denote $P(t,a) = \frac{l(t,a)}{N(t)}$; the probability density function (pdf) $P(t,a)$ describes the distribution of the death rate $a$ throughout the population at $t$ moment. It easily follows from (3.1), (3.2) that the population size $N(t)$ satisfies the equation

$$\frac{dN}{dt} = -N E_t[a] g(N) \qquad (3.3)$$

and the pdf $P(t,a)$ solves the *replicator equation* of the form

$$\frac{dP(t,a)}{dt} = P(t,a)\left(E^t[a] - a\right)g(N). \qquad (3.4)$$

Assume that parameter $a$ has the initial distribution $P(0,a)$; denote

$L_0(\lambda) = \int_A e^{-\lambda a} P(0,a)da$ to be the Laplace transform of the initial distribution.

In order to solve the problem (3.1), let us define formally the auxiliary variable $q(t)$ as the solution to the Cauchy problem

$$\frac{dq}{dt} = g(N), q(0) = 0 . \qquad (3.5)$$

This equation cannot currently be solved because the population size $N(t)$ is unknown. However, the population size and clone densities can be expressed with the help of the variable $q(t)$.

Integrating the equation $\dfrac{dl(t,a)}{dt} = -al(t,a)\dfrac{dq}{dt}$ yields

$$l(t,a) = l(0,a)e^{-aq(t)} = N(0)P(0,a)e^{-aq(t)} . \qquad (3.6)$$

The total population size $N(t)$ is calculated according to Equation (3.2) and is determined by the formula

$$N(t) = N(0)\int_A e^{-aq(t)} P(0,a) da = N(0) L_0(q(t)). \qquad (3.7)$$

The equation for the auxiliary variable $q(t)$ can be written now in a closed form

$$\frac{dq(t)}{dt} = g(N(0) L_0(q(t))), \; q(0) = 0. \qquad (3.8)$$

With the solution to this equation, we can now completely solve the initial problem (3.1), (3.2). The clone densities and population size are given by formulas (3.6), (3.7).

The current parameter distribution $P(t,a)$ is determined by the formula

$$P(t,a) = \frac{l(t,a)}{N(t)} = P(0,a)\frac{e^{-q(t)a}}{L_0(q(t))} . \qquad (3.9)$$

The Laplace transform of the current distribution $P(t,a)$ is given by the formula

$$L_t(\delta) = E^t[e^{-\delta a}] = \int_A \frac{e^{-\delta a - q(t)a}}{L_0(q(t))} P(0,a) da = \frac{L_0(\delta + q(t))}{L_0(q(t))} \qquad (3.10)$$

The mean value of $a$ over time is described by

$$E^t[a] = \int_A \frac{ae^{-q(t)a}}{L_0(q(t))} P(0,a) da = -\frac{d\ln L_0(q)}{dq}\bigg/_{q=q(t)} . \qquad (3.11)$$

Multiplying equation (3.4) by $a$ and integrating yields

$$\frac{dE^t[a]}{dt} = -g(N)Var^t[a]. \tag{3.12}$$

### The power extinction models and inhomogeneous *F*-models

In this section we investigate the second type of inhomogeneous power models assuming that the total size of the inhomogeneous population decreases according to the sub-exponential equation

$$\frac{dN}{dt} = -kN^p, \quad p < 1. \tag{4.1}$$

Let us now construct the inhomogeneous model (3.1)-(3.2) in such a way that its total population size solves equation (4.1). Let $g(N) = kN^{-1}$. Consider inhomogeneous model

$$\frac{dl(t,a)}{dt} = -\frac{kal(t,a)}{N(t)} = -kaP(t,a). \tag{4.2}$$

We will henceforth refer to model (4.2) as a frequency-dependent model, or F-model for brevity.

The total population size of F-model (4.2) solves the equation (see Eq.3.3)

$$\frac{dN}{dt} = -NE^t[a]g(N) = -kE^t[a].$$

So, in order to obtain Equation (4.1), we need to have $E^t[a] = N(t)^p$ for all $t$.

To this end, let us define the auxiliary variable through the equation $\frac{dq}{dt} = g(N) = kN^{-1}$, or in a closed form (see Eq. 3.8)

$$\frac{dq(t)}{dt} = \frac{k}{N(0)L_0(q(t))}, \quad q(0) = 0,$$

where $L_0$ is the Laplace transform of unknown initial distribution. Then (see Eq.3.11)

$$E^t[a] = -\frac{d\ln L_0(q(t))}{dq}.$$

Hence, the equation $E^t[a] = N(t)^p = N(0)^p L_0(q(t))^p$ holds if

$$-\frac{1}{L_0(q(t))}\frac{dL_0(q(t))}{dq} = N(0)^p L_0(q(t))^p,$$

i.e., when

$$\frac{dL_0(q)}{dq} = -N(0)^p L_0(q)^{p+1}.$$

Solving this equation given the initial condition $L_0(0) = 1$, we obtain

$$L_0(q) = (1 + pN(0)^p q)^{-1/p}. \qquad (4.2)$$

It is well known that the function $L_0(\delta) = (1 + \beta\delta)^{-\rho}$ at $\beta > 0$ is the Laplace transform of the Gamma-distribution

$$P(a) = \frac{a^{\rho-1} e^{-\frac{a}{\beta}}}{\beta^\rho \Gamma(\rho)}, \quad a > 0. \qquad (4.3)$$

Hence, the initial distribution of the parameter $a$ should be the Gamma-distribution (4.3) with $\rho = \frac{1}{p}$ and $\beta = pN(0)^p$. This distribution is completely characterized by its mean value $N(0)^p$ and variance $pN(0)^{2p}$.

Now we can compute the auxiliary variable $q(t)$ given the initial distribution (4.3) using the equation:

$$\frac{dq}{dt} = k(N(0)L_0(q))^{-1} = k(N(0))^{-1}(1 + pN(0)^p q)^{\frac{1}{p}}.$$

The solution to this equation under initial condition $q(0) = 0$ is

$$q(t) = \frac{(1-kN(0))^{-1+p}(1-p)t)_+^{-\frac{p}{1-p}} - 1}{pN(0)^p} \qquad \text{for } p \neq 1 \qquad (4.4)$$

and $q(t) = \dfrac{e^{kt}-1}{N(0)}$ for $p=1$.

The current pdf $P(t,a)$ is again the Gamma-distribution, as easily follows from formulas (3.10), (4.2), since

$$L_t(\delta) = \frac{L_0(\delta + q(t))}{L_0(q(t))} = (1 + \frac{pN(0)^p \delta}{1+pN(0)^p q(t)})^{-\frac{1}{p}} = (1+pN(t)^p \delta)^{-1/p} \qquad (4.5)$$

is the Laplace transform of the Gamma-distribution (4.3) with parameters $\rho = \dfrac{1}{p}$ and

$\beta(t) = pN(t)^p$.

The current mean value of the Gamma-distribution with Laplace transform (4.5) is

$$E^t[a] = \rho\beta(t) = N(t)^p. \qquad (4.6)$$

Hence, the total population size of inhomogeneous population (4.2), (4.3) solves the power equation (4.1), as desired.

Next, it follows from formulas (4.4) that $q(t)$ increases monotonically in such a way that $q(t) \to \infty$ as $t \to \infty$ for $p \geq 1$, and $q(t) \to \infty$ as $t \to T$ for $p<1$, where

$$T = \frac{1}{kN(0)^{-1+p}(1-p)} < \infty. \qquad (4.7)$$

The population size for $p \neq 1$ is given by

$$N(t) = N(0)L_0(q(t)) = N(0)\left(1+pN(0)^p q(t)\right)^{-\frac{1}{p}} = N(0)(1+kN(0)^{-1+p}(p-1)t)_+^{\frac{1}{1-p}} \qquad (4.8)$$

and $N(t) = N(0)e^{-kt}$ for $p=1$.

Hence, $N(t)$ decreases monotonically and

$N(t) \to 0$ as $t \to \infty$ for $p \geq 1$ (exponential and super-exponential models);

$N(t) \to 0$ as $t \to T < \infty$ for $p < 1$ (sub-exponential model), where $T$ is given by (4.7).

We see that the life time of a population described by the exponential or super-exponential equation is indefinite; in contrast, a sub-exponential population goes to extinction not asymptotically, as super-exponential power models, but completely disappears at a certain finite time moment. This property makes sub-exponential power models of primary interest.

Coming back to the general power model of extinction (4.1), we want to emphasize that we have found within the frameworks of inhomogeneous population models (3.1)-(3.2) a universal, i.e. for all $p > 0$, *frequency-dependent* representation of power extinction models (*F-model*). Let us summarize our findings in the following Theorem.

**Theorem 2**.

1) *Any power equation*

$$\frac{dN}{dt} = -kN^p, \quad p > 0 \tag{4.9}$$

*describes the dynamics of total size of an inhomogeneous frequency-dependent model*

$$\frac{dl(t,a)}{dt} = -\frac{kal(t,a)}{N(t)} = -kaP(t,a), \tag{4.10}$$

*where* $P(0,a)$, *the initial distribution of the parameter* $a$, *is Gamma-distribution* (4.3) *with* $\rho = 1/p$, $\beta = \beta(0) \equiv pN(0)^p$, *and Laplace transform* $L_0(\delta) = (1 + \beta(0)\delta)^{-1/p}$;

2) *the solution to the power equation exists for all $t > 0$ at $p \geq 1$ and only up to the moment of population extinction, when $N(T) = 0$:*

$$T = \frac{1}{kN(0)^{p-1}(1-p)} \quad \text{for } p < 1. \tag{4.11}$$

3) *the solution to the inhomogeneous F-model (4.10) is*

$$l(t,a) = l(0,a)e^{-aq(t)} \tag{4.12}$$

where the *auxiliary variable* $q(t)$ is given by

$$q(t) = \frac{(1-kN(0))^{-1+p}(1-p)t)_+^{-\frac{p}{1-p}} - 1}{pN(0)^p} \quad \text{when } p \neq 1, \tag{4.13}$$

and by $q(t) = \dfrac{e^{kt}-1}{N(0)}$ for $p=1$;

4) *the total population size at the moment $t$ is*

$$N(t) = N(0)(1 + kN(0)^{-1+p}(p-1)t)_+^{\frac{1}{1-p}} \tag{4.14}$$

for $p \neq 1$ and $N(t) = N(0)e^{-kt}$ for $p=1$;

5) *the current distribution at any time moment $t$ is Gamma-distribution with the mean $E^t[a] = N(t)^p$ and the variance $Var^t[a] = pN(t)^{2p}$.*

**The "internal time" for F-models of extinction**

We have shown that any power model of extinction (4.1) has a canonical representation in the form of F-model (4.10).

Now let us explore the following transformation. Define the following change in time in this model:

$$dq = \frac{kdt}{N(t)}. \tag{5.1}$$

Consider the equation

$$\frac{dx(q,a)}{dq} = -ax(q,a) \tag{5.2}$$

and let

$$l(t,a) = x(q(t),a). \tag{5.3}$$

Then $\dfrac{dl}{dt} = \dfrac{dx}{dq}\dfrac{dq}{dt} = -\dfrac{ax(q(t),a)k}{N(t)} = -\dfrac{akl(t,a)}{N(t)}$.

Hence, $l(t,a)$, as defined by (5.3), solves equation (4.10). Noticeably, through change of time $t \to q(t)$ as defined in (5.1), the F-model (4.10) becomes reduced to a simple Malthusian model of extinction (5.2). An explicit expression for $q(t)$ is given by (4.13).

The possibility of transformation of the initial power model (4.1) to the form (5.2) reveals some interesting properties of the power models of extinction.

We may interpret the variable $q(t)$ as a natural "internal" time for the F-models. Such interpretation becomes possible because *each clone $l(q,a)$ with respect to this time scale evolves as if it does not depend on other clones and on the population as a whole*. The F-model with respect to the internal time $q$ defined by Equation (5.1) becomes identical to inhomogeneous Malthusian model with respect to the "common", or "external" time $t$, which describes free development of all clones in the population. Notice also that $q(t)$ is the only time scale for the Equation (4.10), which possesses by this property.

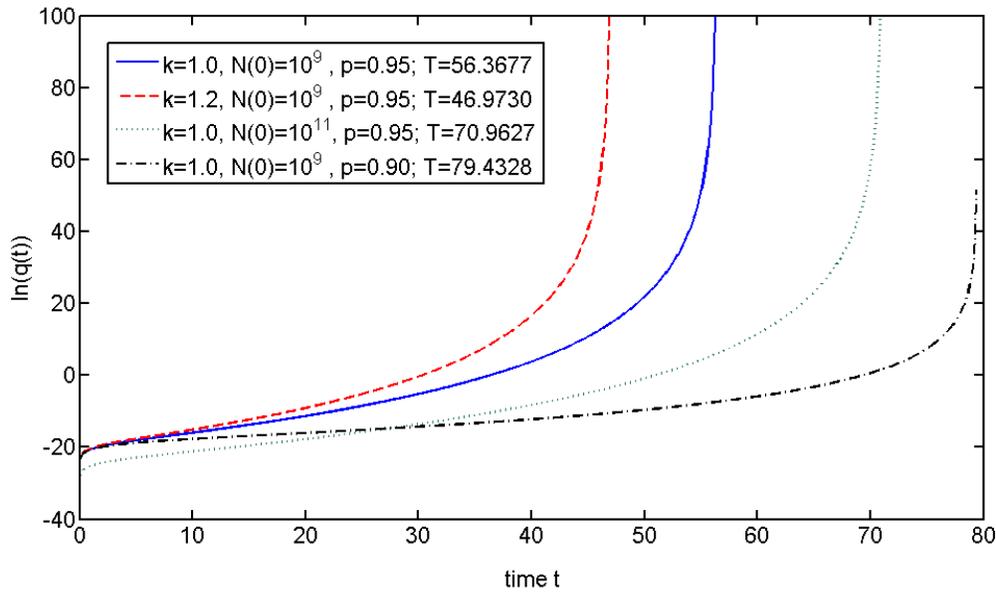

**Figure 5.** Plot of internal time $q(t)$, defined in Equation (4.13), for $p<1$, in logarithmic scale, against real time.

It follows from the explicit expression of the internal time (4.13) that if $p \geq 1$, then $q(t)$ is finite for all $t$ and $q(t) \to \infty$ as $t \to \infty$. In contrast, if $p<1$, then $q(t) \to \infty$ as $t \to T < \infty$, where

$$T = \frac{N(0)^{1-p}}{k(1-p)}. \tag{5.5}$$

This means that a finite "real time" interval $(0,T)$ corresponds to infinite duration of "internal" time. Every clone $l(t,a)$, as well as the population as a whole, tend to extinction simultaneously at the moment $T$. This is in contrast with the model considered in previous sections, where every clone has its own life duration.

It is easy to show, that the value of $T$ at a given $N(0)$ is minimal when

$$p = p(N(0)) \equiv p_{opt} = 1 - \frac{1}{lnN(0)}. \tag{5.6}$$

Indeed, $\dfrac{dT}{dp} = \dfrac{N(0)^{1-p}}{k(1-p)}(-lnN(0) + \dfrac{1}{1-p})$, so $\dfrac{dT}{dp} = 0$ if $p = 1 - \dfrac{1}{lnN(0)}$. In order to guarantee that $0 < p < 1$ we should assume that $lnN(0) > 0$. The minimal value of $T$ is attained when $p = p_{opt}$ and is equal to

$$T_{min} = \frac{lnN(0) N(0)^{\frac{1}{lnN(0)}}}{k} = \frac{e}{k} lnN(0) \tag{5.7}$$

since $N^{\frac{1}{lnN}} \equiv e.$

Summarizing the obtained results leads to the following Proposition, where we assume that $lnN(0) > 0$.

**Proposition 1.** *The life duration of sub-exponential population* (4.10) *with any* $0 < p < 1$ *is bounded from below by* $T \geq T_{min} = \frac{e}{k} lnN(0)$. *For a given* $N(0)$, *this boundary is attained if* $p = p_{opt} = 1 - \frac{1}{lnN(0)}$. *Conversely,* $T = T_{min}$ *at a given* $p$ *if the initial population size is equal to*

$$N(0) = \exp(\frac{1}{1-p}).$$

Remark that for any model of extinction of inhomogeneous population of the form (3.1)-(3.2), the internal time can be defined by the equation $\frac{dq}{dt} = g(N)$. The model (3.1)-(3.2) with respect to the internal time $q$ becomes identical to the inhomogeneous Malthusian model. The properties of the internal time $q(t)$ critically depend on the function $g$ and on the initial distribution of the Malthusian parameter $a$.

### Dynamical principles of minimum of Shannon information loss

The Principle of Minimum of Cross-Entropy (MinxEnt) (see Kullback 1968, Beck 2009) is based on the hypothesis that subject to precisely stated prior data, the probability distribution that best represents the current state of knowledge is the distribution with the minimal cross-entropy, known also as KL-divergence between the current distribution *m* and a reference distribution *r*. Recall the definition of KL-divergence (Kullback 1959), which is a generalization of the Boltzmann-Gibbs entropy or the Shannon information:

$$I_{KL}[m:r] = \int_A m(x) \ln \frac{m(x)}{r(x)} dx. \qquad (6.1)$$

The KL-divergence can be considered as a sort of "distance" (but not a metric) between two distributions, *m* and *r*. In our model the value of $I_{KL}[P_t : P_0]$, where $P_t = P(t,a)$ is defined by (3.9), can be computed using the formula

$$I_{KL}[P_t:P_0] = \int_A P(t,a)(-aq(t) - \ln L_0(q(t)))da = -q(t)E^t[a] - \ln L_0(q(t)) \quad (6.2)$$

The mean value $E^t[a] = -\dfrac{d}{dq}\ln L_0(q(t))$ (see 1.8), therefore,

$$I_{KL}[P_t:P_0] = q(t)\dfrac{d}{dq}\ln L_0(q(t)) - \ln L_0(q(t)). \quad (6.3)$$

Frequencies of the clones (3.9) with respect to the internal time are given by

$$P(q,a) = P(0,a)\dfrac{e^{-aq}}{L_0(q)}. \quad (6.4)$$

Then

$$I_{KL}[P_q:P_0] = q\dfrac{d}{dq}\ln L_0(q) - \ln L_0(q) \text{ according to (6.3)}.$$

Figures 6 and 7 show the dramatic difference in behaviors of $I_{KL}[P_q:P_0]$ vs. the internal time $q$ and $I_{KL}[P_t:P_0]$ vs. the real time $t$; in both cases $P_0(a)$ is the Gamma-distribution given by (4.3) with $\rho = 1/p$ and $\beta = pN(0)^p$; $p = 0.96, N(0) = 10^{11}$.

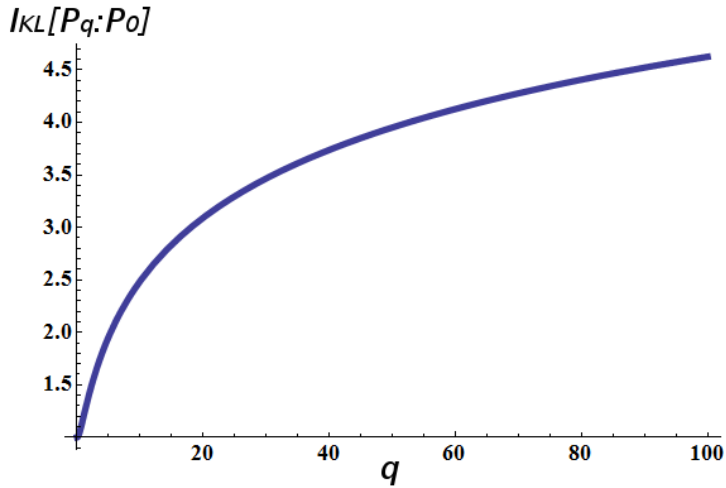

**Figure 6.** The plot of $I_{KL}[P_q:P_0]$, defined in Equation (6.3), plotted against internal time $q$.

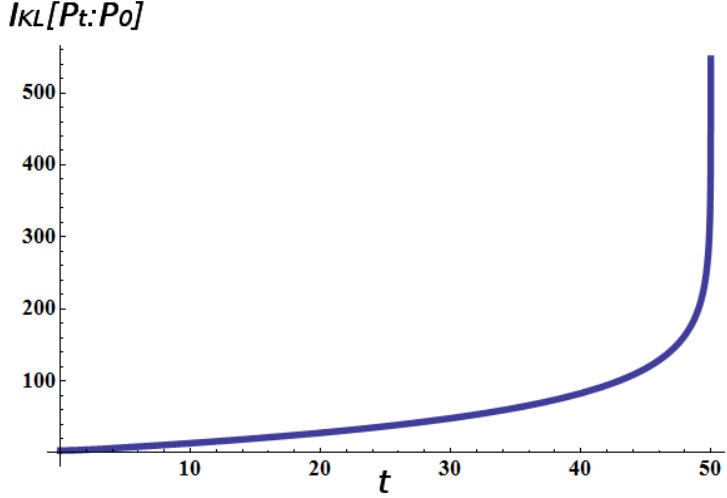

**Figure 7.** The plots of $I_{KL}[P_t:P_0]$, defined in Equation (6.3), plotted against "real" time $t$.

An interesting property of the $I_{KL}[P_t:P_0]$ dynamics is given by the following Proposition.

**Proposition 2.**

$$\frac{d}{dt}I_{KL}[P_t:P_0] = \frac{q(t)}{N(t)}Var^t[a] \ . \quad (6.5)$$

Indeed, according to equation (5.6),

$$\frac{dI_{KL}[P_t:P_0]}{dt} = \frac{d}{dt}\left(-q(t)E^t[a] - lnL_0(q(t))\right) =$$
$$= -q(t)\frac{dE^t[a]}{dt} - E^t[a]\frac{dq}{dt} - (\frac{d}{dq}lnL_0(q(t)))\frac{dq}{dt} =$$
$$= \frac{q(t)Var^t[t]}{N(t)},$$

since $\frac{d}{dq}lnL_0(q(t)) = -E^t[a]$ and $\frac{dE^t[a]}{dt} = -\frac{Var^t[a]}{N(t)}$ according to (3.11) and (3.12) . Q.E.D.

Therefore, the KL-divergence increases monotonically, as we have seen in Figures 5-7, and the rate of its increase is proportional both to the current variance of the parameter *a* and the value of the internal time $q(t)$.

The importance of the KL-divergence $I_{KL}[P_q : P_0]$ is two-fold. Firstly, $I_{KL}[P_q : P_0]$ is a general characteristic of current state of a population that goes to extinction, which shows the "distance" between the current distribution $P_q$ and the initial distribution $P_0$. Secondly, the KL-divergence plays a central role in the Principle of minimum of relative Boltzmann-Gibbs entropy, MinxEnt. According to known results (see, e.g., Borland et al, 1998, Dukkipati et al., 2006, Tsallis 2008, etc.), the distribution $m(x)$ that provides minimum for $I_{KL}[m:r]$ given the mean value of *m* equal to *s*, is the Boltzmann distribution

$$m(x) = \frac{\exp(-sx)}{Z} r(x) , \quad (6.6)$$

where Z is the normalization factor (the "partition function").

It is easy to see that distribution $P_q = P(q,a)$ (6.4), which is defined by the solution to inhomogeneous model (4.12), coincides up to notation with the Boltzmann distribution (6.6) that provides the minimal value of KL-divergence $I_{KL}[P_q : P_0]$.

As before, we submit that the information measure for dynamical models and systems should be chosen in accordance with the system dynamics. In the case of F-models of population decrease, the distribution of individual frequencies is the Boltzmann distribution at each time moment, and accordingly, the KL-divergence, which is a generalization of classical Shannon information, is the appropriate information measure for the model. The value of KL-divergence $I_{KL}[P_q : P_0]$ can be interpreted as *Shannon information loss* during the internal time interval [0, *q*] in a decreasing population.

Therefore, the dynamics of F-model (4.10) is such that at each moment of the *internal time,* the Shannon information loss is minimal. It means that the Principle of minimum of relative entropy, MinxEnt, in the form of the principle of minimum of Shannon information loss for F- model, holds automatically as a result of model dynamics.

Let us emphasize that we do not seek an unknown distribution that would minimize the KL-divergence $I_{KL}[P_q : P_0]$ subject to a particular set of constraints, in contrast to the common approach. Instead, we have the solution (4.12), (4.13) of F-model (4.10), which produces distribution (6.4) at each time moment. With this distribution, we can compute at each moment of internal time *q* the mean of the death rate, knowing only the initial distribution. Hence, *given the initial distribution of clone frequencies in the population, the knowledge of only the mean value of the death rates at any moment of internal time q yields complete knowledge of the population distribution at that moment due to the principle of minimum of KL-divergence*.

This means that the dynamics of F-models is governed by the Principle of minimum of Shannon information loss.

**Application of the model to time perception**

We may surmise that the aforementioned formally defined internal time, in application to one of the largest biological systems, the human brain, can be interpreted as a type of an internal clock model. In this section we will first provide background on the currently accepted biological models of time perception, and will then provide a possible application of one of the proposed extinction models to understanding alterations in time perception in a dying brain.

*Background information on time perception*

As can be seen through a survey of literature, subjective sense of time can be affected temporarily or permanently due to changes in synchronization between an "internal clock" and external "subjective" time. Arguably the most influential internal clock model, based on scalar expectancy theory (SET), was proposed by Gibbon (1977; Gibbon et al. 1984). According to SET, temporal processing is regulated by a pacemaker, switch and accumulator. The switch, controlled by attention, regulates the number of pacemaker pulses (clock ticks) that are collected into the accumulator. Time estimates depend on a number of "pulses" that have been accumulated between the switches: the more pulses have accumulated, the longer the perceived time interval.

This model has since been modified and expanded, and the currently accepted most plausible model of time perception is the striatal beat frequency mode (SBF), which has been proposed by Matell and Meck (2000; 2004). In the SBF, timing is based on the coincidental activation of medium spiny neurons by cortical neural oscillators in the basal ganglia in the midbrain. A broad array of studies have shown that injections of substances that act as dopamine antagonists, such as cocaine and metamphetamine, cause shorter tasks to be perceived as long, slowing down the perception of time, while drugs that activate dopamine receptors, such as haloperidol and pimozide, speed up the perception of time (Meck 1996; 2005; Coull 2011). Moreover, patients with Parkinson's disease, which involves degeneration of dopaminergic substances in the basal ganglia, and specifically in substantia nigra par compacta (SNc), exhibit impaired timing perception (Malapani et al. 1998; Rammsayer et al. 1997). Similar problems with time perception have also been observed in patients with schizophrenia (Davalos et al. 2002; Penney et al. 2005) and attention deficit hyperactivity disorder (Levy and Swanson 2001; Barkley et al. 2001). These studies suggest that there exists an "optimal" level of dopamine that aligns one's internal clock with external time.

Noticeably, temporary changes to time perception can be affected by emotions (Droit-Volet and Meck 2007; Droit-Volet 2013) and reactions to threatening situations, where involvement of the amygdala in emotional memory contributes to creation of secondary encoding of memories. This can create an erroneous a sense of events spanning a greater period of time than has actually passed and appears to be a function of recollection, not perception of time (Stetson et al. 2007).

Another particularly interesting case of alterations in time perception could be occurring during near-death experiences (French 2005). Anecdotal evidence suggests that time intervals during a near-death experience are perceived to be much longer than they actually are. Moreover, in 2013, Borjigin et al. (2013) reported a high frequency neurophysiological activity in the near-death state of rats undergoing experimentally induced cardiac arrest. The levels of neurophysiological activity in the near-death state exceeded those found during conscious waking state, suggesting that the brain in fact might be highly active in the near-death state.

These observations, coupled with the current state of knowledge about changes time perception, allow formulation of the following hypothesis: During a near-death experience, lack of oxygenation and nutrient access eventually causes death of all brain cells, including the dopaminergic substances in the basal ganglia, causing an experience of internal time to become increasingly longer compared to external time. This could be a possible mechanism to account for the perception of time that allows "life to flash before one's eyes" in the moments preceding death.

*Application of the proposed model to understanding time perception in a dying brain*

The second model of population extinction proposed in this paper allows decoupling of internal and external time. The proposed interpretation of obtained results allows simulating internal time tending to infinity in finite time interval, recapitulating qualitatively the possible experience of time perception in a dying brain.

Recall that in this model we assume that the population (e.g., of neural cells) is composed of clones $l(t,a)$, which here stands for a set of all individuals in the population having the death rate (Malthusian parameter) equal to $a$.

According to Theorem 2, the power equation (4.1) with $p > 0$ describes the dynamics of total size of inhomogeneous frequency-dependent extinction model $\frac{dl(t,a)}{dt} = -kaP(t,a)$, where the initial distribution of the parameter $a$, $P(0,a)$, is the Gamma-distribution. Populations described by sub-exponential population with $p < 1$ go to extinction at certain finite time moment, i.e. $N(T) = 0$, when $T = \frac{N(0)^{1-p}}{(k(1-p))}$. Within the context of our model, this suggests that the time to $T$ depends on either the initial number of neurons, or connections between them, both of which are logically consistent.

Making a formal change of time $t \to q$ through equation $dq = \frac{k}{N(t)} dt$ allows rewriting the equation for the dynamics of cell clones $l(t,a)$ with respect to this new independent variable $q$ as $\frac{dl(q,a)}{dq} = -al(q,a)$, which becomes a standard Malthusian equation of population extinction. Here we propose that $q(t)$ is a possible representation of internal time. It is important that if $0 < p < 1$, then $q(t) \to \infty$ as $t \to T < \infty$, where $T$ is the time moment of population extinction. The internal time $q(t)$ as a function of "real time" $t$ is given by (4.13); the plot of internal time $q(t)$ for $p < 1$ against real time is given in Figure 5.

There are three parameters that determine the dynamics of the internal time $q(t)$: $k$, $N(0)$ and $p$. As one can see, larger values of $k$, which defines the death rate of the population as a whole, decrease extinction time $T$; increasing the initial value of $N(0)$ predictably delays extinction time $T$; dependence of $T$ on the value of $p$ is non-monotonic. $N$ can be interpreted as either the number of neural cells, or possibly the number of connections between cells. The latter interpretation might be preferable, since demise of connections between cells in a damaged or dying brain can lead to different parts of the brain living and dying as if separate from each other, which underlies the assertion that $q(t)$ can in fact be representing internal time. An average human brain is estimated to range between 85 and 120 billion neurons (Herculano-Houzel 2009), so we can estimate $N(0)$ to be roughly $10^{11}$. The number of synapses (based on 1000 per neuron estimate) is about $10^{14}$, or 100 trillion.

Furthermore, from Equation (4.13), the expression $kN(0)^{(p-1)}$ can be seen as a scaling quantity for "external" time.

For the power $p$, we do not yet have a well-established biological interpretation but we can identify certain properties. Specifically, if $p > 1$, the population defined in power Equation (4.1) lives forever, but for $p < 1$ it will go extinct in finite time, defined by $T$. Hence, the parameter $p$ can be estimated based on the duration of the process of population extinction. Equation (5.6) and the Proposition 1 above give a simple estimation of $p$ with respect to the initial population size $N(0)$.

The inhomogeneous *F*-model (4.10) noticeably does not depend on $p$; the rate of extinction of cells, or synaptic connections between them, is proportional to their frequency. However, the distribution of parameter *a*, which defines population clones, does depend on $p$. Specifically, the expected value and the variance of parameter $a$, which define completely the initial Gamma-distribution, both depend on the value of $p$. Therefore theoretically, the mean value of the parameter $a$ can be used for estimation of the power $p$, as given by $E^t[a] = N(t)^p$.

According to Proposition 1, there exists a single value of $p = p_{opt}$, given by equation (5.6), which minimizes $T$; for instance, $p_{opt} = 0.96$ for $N(0) = 10^{11}$ and $p_{opt} = 0.97$ for $N(0) = 10^{14}$; see Figure 8.

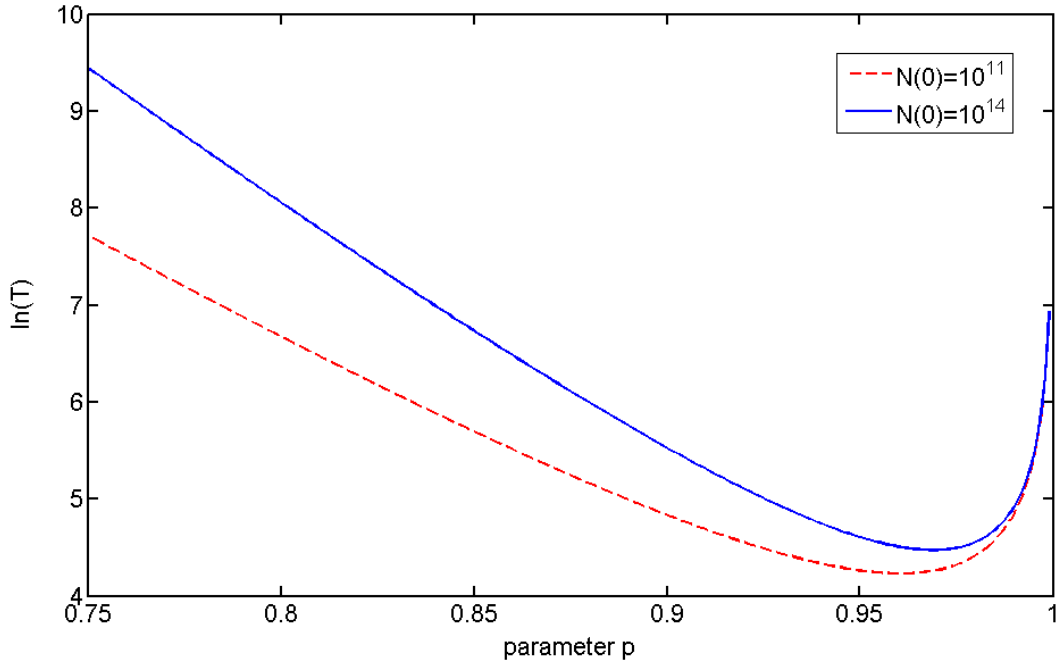

**Figure 8.** The plot of extinction moment $T$, defined in Equation (4.11), plotted in logarithmic scale against the power $p$ for $N(0) = 10^{11}$ (dotted) and $N(0) = 10^{14}$ (solid).

Furthermore, by plotting $p_{opt}$ vs $N(0)$ according to Equation (5.6), we can see that for a very large magnitude of values of $N(0)$, $10^{10} < N(0) < 10^{15}$, the values of $p_{opt}$, which provide minimum of $T$ at given initial population size, are $0.957 < p_{opt} < 0.971$, demonstrating that the "optimal" value of $p$ belongs to a very narrow range.

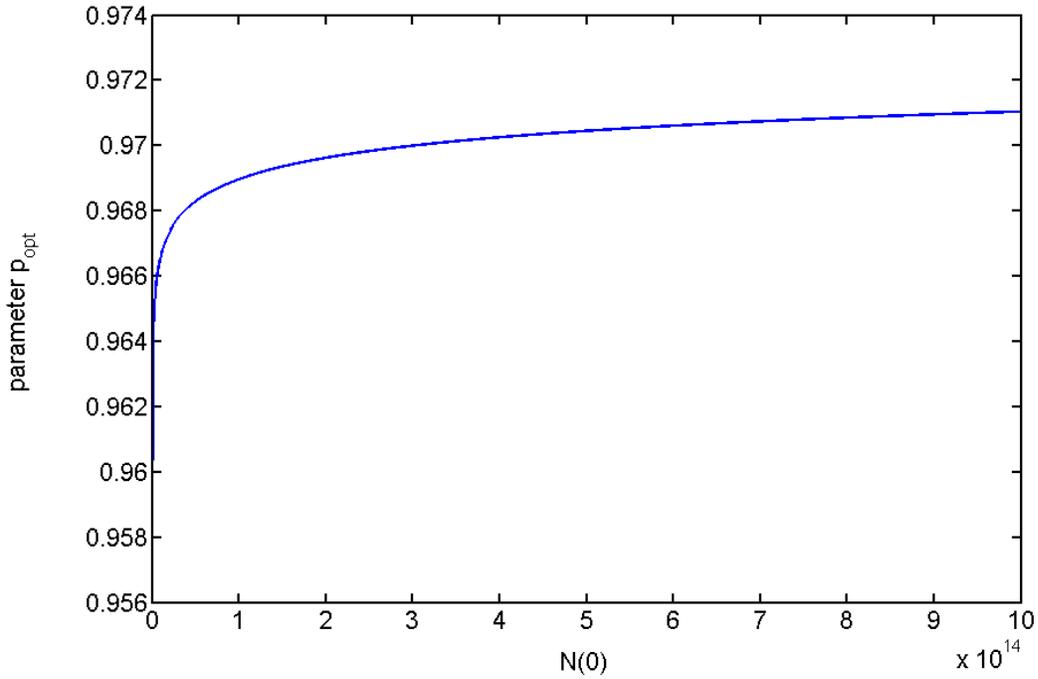

**Figure 9.** The plot of the optimal value $p_{opt}$ given by Equation (5.6), against $N(0)$, plotted for $10^{10} < N(0) < 10^{15}$.

The minimal value of $T$ when $p = p_{min}$ is given by Proposition 1 and is proportional to $lnN(0)$. The duration of death of the brain core in the state of clinical death is between 5 and 20 minutes (Safar 1988). According to Equation (5.7), $T = 300\,sec$ if $\frac{lnN(0)}{k} = \frac{300}{e} \cong 110$, i.e. if for instance $N(0) = 10^{11}$ and $k = 0.23$, or if $N(0) = 10^{14}$ and $k = 0.29$.

Then, within the framework of our model, if the process of death of brain core can be described by power equation (4.1), $\frac{dN}{dt} = -kN^p$, with $10^{11} < N(0) < 10^{14}, 0.96 < p < 0.97$, and $0.23 < k < 0.29$, then during $T = 5$ min of "external", or chronological time, the "internal" time of the underlying *F*-model can become indefinitely large, and potentially infinite.

**Discussion**

In this paper we study two inhomogeneous models of population extinction. In the first model we assume that the population is composed of independent clones, each of which decreases according to the sub-exponential equation. In the second model we assume that the total size of the inhomogeneous population decreases according to the sub-exponential equation; we reveal and study the underlying inhomogeneous frequency-dependent population model. These conceptual models can be potentially applied for investigation of extinction processes in different ecological and biological systems. Furthermore, any individual and even any single organ, e.g., a human brain, can be viewed as a complex community of large populations of cells and connections between them that undergo extinction with the death of the individual and could thus potentially be described using these models.

We have shown that the dynamics of clone frequencies in both models follows the Principle of minimum of information loss, but the adequate information measures are different for these models. We submit that the information measure for dynamical models and systems should be chosen in accordance with the system dynamics. Then, the adequate information measure for the distribution of individual frequencies in the 1st model is Tsallis relative $q$-entropy, while the adequate information measure in the 2nd model is the classical Boltzmann-Gibbs relative entropy (or Shannon information loss). The Tsallis entropy and distribution include the standard Shannon entropy and the Boltzmann-Gibbs distribution as a special case when $q \to 1$.

A novel and important mathematical result that comes from this investigation is the following. The Principle of minimum of relative entropy, MinxEnt, was successfully applied to various statistical, physical and biological problems as a method for inference of unknown distribution, subject to some given constraints (see Kapur, Kesavan 1992). For example, Jaynes (1957) and followers have shown that essentially all known statistical mechanics can be derived from this principle. What is important that in those models this principle formally is a hypothesis, and it is not clear why one can expect it to work as a description of nature. This problem has been intensively discussed in the literature over the last 50 years.

In this paper we show that for considered models of population extinction, the MinxEnt in the form of the Principle of minimum of information loss is neither an inference algorithm nor an external principle but a mathematical assertion that can be derived from the system dynamics instead of being postulated. With the solution to the extinction models, we can compute the current mean values of the death rates at any instant. Then, treating these mean values as constraints, we can show that the Tsallis and Boltzmann distributions that minimize the Tsallis or Shannon information loss correspondingly coincide with the solutions to the extinction models, which were obtained independently of the MinxEnt algorithm. Hence, the principle of the minimum of information loss can be considered as the variation principle that governs the dynamics of population extinction.

Let us remind that the "information loss" here means the divergence in distributions of death rates between population at the initial time point and population at the current time moments. One can even further speculate that loss of neural synapses and neural cells during brain death, in the event that it occurs in accordance with the proposed model, would result in similar minimization of loss of qualitative information, such as knowledge and memories. This hypothesis would of course require further investigation.

A fundamental property of the Tsallis entropy is that it is non-additive for independent subsystems. Thus, in the 1st model, for the system of sub-exponentially decreasing clones the information about two exhaustive independent subsystems is insufficient to obtain the information about system as a whole. This "non-reductionist" character of the system might reflect the fact that different clones that make up the population go to extinction at different time moments. In contrast, in the 2nd model, all clones and the population as a whole go to extinction simultaneously, at the same moment of real "chronological", or "external" time. Additionally, there exists an "internal time" in the population as a whole, such that with respect to this internal time the system is described by the standard Malthusian model of extinction.

A possible and perhaps unexpected application of the proposed modeling framework lies in providing a possible explanation underlying time perception. Biologically, perception of time and alignment of one's "internal clock" with "external" chronological time appears to be primarily affected by the activity of the dopamine in basal ganglia in the midbrain (Mattell and Meck 2000;

2004). We hypothesize that when a brain is dying, such as after a cardiac arrest, the eventual cessation of activity of all neural cells, including dopamine system in the basal ganglia, can cause a perception of time to increase dramatically compared to "external" chronological time.

The proposed mathematical model allows replicating this effect through decoupling "internal" and "external" time using a parametrically heterogeneous sub-exponential power equation, where cells, or possibly synapses $N(t)$, in the population die at different rates. We divided the population of neural cells into clones, and death rate of each clone is proportional to its frequency in the population.

We derived an equation for $q(t)$, a variable that can be interpreted as describing "internal time", because each clone, or subpopulation, with respect to this time scale evolves as if it does not depend on other clones or on the population as a whole. Within a context of a dying brain, this would describe a situation when loss of connections between cells in the dying brain would indeed lead to subpopulations of cells dying independently of each other, which is logically consistent. Furthermore, from this equation we were able to also specify time of death $T$, where population of cells goes to extinction, i.e., when $N(T) = 0$. Conversely, knowing the specific time of death $T$, we can estimate other model parameters.

One can also apply this model to explore hypotheses about what may be happening with time perception in the moments preceding death. Given that the malfunction of the dopamine system in the basal ganglia is responsible for alterations in time perception (Mattell and Meck 2000; 2004), then within the frameworks of this model, death of cells, or synapses $N(t)$ will result in internal time $q(t) \to \infty$ in a finite period of time, denoted by $T$. The details of what happens with time perception prior to brain death are not yet understood, including whether internal time indeed increases in the moments preceding death, allowing one to experience more than would normally be possible (Bierce 2008).

It is of course not known exactly what mechanisms could be governing perception of time in the human brain. Here we proposed a possible mathematical formalization, which allows making logically consistent predictions. We have identified a small number of key parameters that could be involved in the decoupling of "internal" and "external" time in the human brain, and proposed an explanation for mechanisms underlying time perception. We hope that this conceptual model can lay a foundation for further mathematical and theoretical exploration of this complex topic, which eventually might yield results to deepen the understanding of diseases, such as Parkinson's, schizophrenia and ADHD, where accurate time perception is compromised.

**Conclusions**

Here we have investigated the properties of power models of population extinction. It is our hope that these results may find application in a variety of fields, including ecology, paleontology, species extinction, as well as possibly bring insights into the dynamics that govern such complex systems as cancer or even the human brain.

**Acknowledgements**

This research was partially supported by the Intramural Research Program of the NCBI, NIH. The authors would also like to thank the anonymous reviewers, whose comments and suggestions contributed to significant improvement of the manuscript.

**References**

1. Barkley, RA Murphy, K. R., & Bush, T. (2001). Time perception and reproduction in young adults with attention deficit hyperactivity disorder. *Neuropsychology*, *15*(3), 351.

2. Beck, C. (2009). Generalized information and entropy measures in physics. *Contemporary Physics*, *50*(4), 495-510.

3. Bierce, A. (2008). An occurrence at Owl Creek Bridge and other stories. Courier Corporation.

4. Borjigin, J., Lee, U., Liu, T., Pal, D., Huff, S., Klarr, D., & Mashour, G. A. (2013). Surge of neurophysiological coherence and connectivity in the dying brain. *Proceedings of the National Academy of Sciences*, *110*(35), 14432-14437.

5. Borland, L., Plastino, A. R., & Tsallis, C. (1998). Information gain within nonextensive thermostatistics. *Journal of Mathematical Physics*, *39*(12), 6490-6501.

6. Coull, J. T., Cheng, R. K., & Meck, W. H. (2011). Neuroanatomical and neurochemical substrates of timing. *Neuropsychopharmacology*, *36*(1), 3-25.

7. Cressie, N., & Read, T. R. (1984). Multinomial goodness-of-fit tests. *Journal of the Royal Statistical Society. Series B (Methodological)*, 440-464.

8. Davalos, D. B., Kisley, M. A., & Ross, R. G. (2002). Deficits in auditory and visual temporal perception in schizophrenia. *Cognitive Neuropsychiatry*, *7*(4), 273-282.

9. Dodson, C. T. J. (2008). On the entropy flows to disorder. *arXiv preprint math-ph/ 0811.4318*. http://www.maths.manchester.ac.uk/ kd/PREPRINTS/ GammaEntropyFlows.pdf

10. Droit-Volet, S., & Meck, W. H. (2007). How emotions colour our perception of time. *Trends in cognitive sciences*, *11*(12), 504-513.

11. Droit-Volet, S. (2013). Time perception, emotions and mood disorders. *Journal of Physiology-Paris*, *107*(4), 255-264.

12. Dukkipati, A., Murty, M. N., & Bhatnagar, S. (2006). Nonextensive triangle equality and other properties of Tsallis relative-entropy minimization. *Physica A: Statistical Mechanics and its Applications*, *361*(1), 124-138.

13. Von Foerster, H., Mora, P. M., & Amiot, L. W. (1960). Doomsday: Friday, 13 November, ad 2026. *Science*, *132*(3436), 1291-1295.

14. French, C. C. (2005). Near-death experiences in cardiac arrest survivors. *Progress in Brain Research*, *150*, 351-367.

15. Gibbon, J. (1977). Scalar expectancy theory and Weber's law in animal timing. *Psychological review*, *84*(3), 279.

16. Gibbon, J., Church, R. M., & Meck, W. H. (1984). Scalar timing in memory. *Annals of the New York Academy of sciences*, *423*(1), 52-77.

17. Gorban, A., Gorban, P., and Judge, G. (2010). Entropy: The Markov Ordering Approach. *Entropy*, *12*, 1145-1193.

18. Herculano-Houzel, S. (2009). The human brain in numbers: a linearly scaled-up primate brain. *Frontiers in human neuroscience*, *3*.

19. Jaynes, T. (1957). Information theory and statistical mechanics 1. *Phys Rev*, 106, 620–630.

20. Johansen, A., & Sornette, D. (2001). Finite-time singularity in the dynamics of the world population, economic and financial indices. *Physica A: Statistical Mechanics and its Applications*, *294*(3), 465-502.

21. Kapitza, S. P. (1996). The phenomenological theory of world population growth. *Physics-uspekhi*, *39*(1), 57.

22. Kapur, J.N., Kesavan, H.K. (1992). *Entropy Optimization Principles with Applications.* San Diego, Academic Press.

23. Karev, G. P. (2003). Inhomogeneous models of tree stand self-thinning. *Ecological Modelling*, *160*(1), 23-37.

24. Karev, G. P. (2005). Dynamics of inhomogeneous populations and global demography models. *Journal of Biological Systems*, *13*(01), 83-104.

25. Karev, G. P., & Koonin, E. V. (2013). Parabolic replicator dynamics and the principle of minimum Tsallis information gain. *Biol. Direct*, *8*, 8-19.

26. Karev, G. P. (2014). Non-linearity and heterogeneity in modeling of population dynamics. *Mathematical biosciences*, *258*, 85-92.

27. Karev, G. P. (2010). On mathematical theory of selection: continuous time population dynamics. *Journal of mathematical biology*, *60*(1), 107-129.

28. Karev, G. P. (2010). Replicator equations and the principle of minimal production of information. *Bulletin of mathematical biology*, *72*(5), 1124-1142.

29. Kullback, S. (1968). *Information theory and statistics*. Courier Corporation.

30. Levy, F., & Swanson, J. M. (2001). Timing, space and ADHD: the dopamine theory revisited. *Australian and New Zealand Journal of Psychiatry*, *35*(4), 504-511.

31. Malapani, C., Rakitin, B., Levy, R. S., Meck, W. H., Deweer, B., Dubois, B., & Gibbon, J. (1998). Coupled temporal memories in Parkinson's disease: a dopamine-related dysfunction. *Cognitive Neuroscience, Journal of*, *10*(3), 316-331.

32. Matell, M. S., & Meck, W. H. (2000). Neuropsychological mechanisms of interval timing behavior. *Bioessays*, *22*(1), 94-103.

33. Matell, M. S., & Meck, W. H. (2004). Cortico-striatal circuits and interval timing: coincidence detection of oscillatory processes. *Cognitive brain research*, *21*(2), 139-170.

34. Meck, W. H. (1996). Neuropharmacology of timing and time perception. *Cognitive brain research*, *3*(3), 227-242.

35. Meck, W. H. (2005). Neuropsychology of timing and time perception. *Brain and cognition*, *58*(1), 1-8.

36. Penney, T. B., Meck, W. H., Roberts, S. A., Gibbon, J., & Erlenmeyer-Kimling, L. (2005). Interval-timing deficits in individuals at high risk for schizophrenia. *Brain and cognition*, *58*(1), 109-118.

37. Rammsayer, T., & Classen, W. (1997). Impaired temporal discrimination in Parkinson's disease: temporal processing of brief durations as an indicator of degeneration of dopaminergic neurons in the basal ganglia. *International Journal of Neuroscience*, *91*(1-2), 45-55.

38. Read, T. R., & Cressie, N. A. (1988). *Goodness-of-fit statistics for discrete multivariate data*. Springer, New York, NY, USA.

39. Safar, P. (1988). Resuscitation from clinical death: pathophysiologic limits and therapeutic potentials. *Critical care medicine*, *16*(10), 923-941.

40. Stetson, C., Fiesta, M. P., & Eagleman, D. M. (2007). Does time really slow down during a frightening event. *PLoS One*, *2*(12), e1295.
41. Szathmáry, E., & Smith, J. M. (1997). From replicators to reproducers: the first major transitions leading to life. *Journal of theoretical biology*, *187*(4), 555-571.
42. Tsallis, C. (2009). *Introduction to nonextensive statistical mechanics* (pp. 37-43). New York: Springer.
43. von Kiedrowski, G. (1993). Minimal replicator theory I: Parabolic versus exponential growth. In *Bioorganic chemistry frontiers,* pp. 113-146. Springer Berlin Heidelberg.